\documentclass{article}

\usepackage{PRIMEarxiv}
\usepackage{float}
\usepackage[utf8]{inputenc} 
\usepackage[T1]{fontenc}    
\usepackage{hyperref}       
\usepackage{url}            
\usepackage{booktabs}       
\usepackage{amsfonts}       
\usepackage{nicefrac}       
\usepackage{microtype}      
\usepackage{lipsum}
\usepackage{fancyhdr}       
\usepackage{graphicx}       
\graphicspath{{media/}}     

\title{Reporting Risks in AI-based Assistive Technology Research:\\A Systematic Review
}

\author{
  ZAHRA AHMADI, PETER R. LEWIS \\
  Ontario Tech University \\
  Oshawa, Canada\\
  \texttt{zahra.ahmadi@ontariotechu.net}\\ 
  \texttt{peter.lewis@ontariotechu.ca} \\
   \And
  MAHADEO A. SUKHAI \\
  The Canadian National Institute for the Blind (CNIB) \\
  Toronto, Canada\\
  Ontario Tech University \\
  Oshawa, Canada\\
  \texttt{Mahadeo.Sukhai@cnib.ca} \\
}

\begin{document}
\maketitle

\begin{abstract}
Artificial Intelligence (AI) is increasingly employed to enhance assistive technologies, yet it can fail in various ways. We conducted a systematic literature review of research into AI-based assistive technology for persons with visual impairments. Our study shows that most proposed technologies with a testable prototype have not been evaluated in a human study with members of the sight-loss community. Furthermore, many studies did not consider or report failure cases or possible risks. These findings highlight the importance of inclusive system evaluations and the necessity of standardizing methods for presenting and analyzing failure cases and threats when developing AI-based assistive technologies.
\end{abstract}


\section{Introduction}
\textsl{Assistive technology} is an umbrella term implying a wide range of tools, devices, services, and systems to increase the functionality and independence of their users~\cite{whoassistive2023}. Persons with disabilities widely use these tools, which may be based on Artificial Intelligence (AI) for various purposes, like virtual assistants for navigation, image description, and recommendation. AI systems can fail unexpectedly, and these cases are widely reported and discussed,~\cite{mcgregorpreventing2021, williamsunderstanding2021}. However, there are few reports on the failures of AI-based assistive technologies. One explanation for this is the failures and safety of these assistive technologies are not included in the scope of work when doing research or making a new assistive technology. 
By not considering failures in assistive technology, the potential risks associated with them will stay out of our sight.

The consequences of this ignorance could lead to failures that cause actual harm to users. Moreover, not being appropriately informed about a product's associated risks and possible failures could make it more or less likely for persons with disabilities to trust and use such a product. For example, depending on the user's attitude, they may mistrust it because the product sounds completely safe when, in fact, there are important exceptions. Alternatively, they may not trust it because there needs to be more information about the situations in which the product may fail when these will never be encountered for practical purposes. So, the opportunity to make a well-informed decision is missed.

This work presents a systematic literature review of the existing research that introduces an AI-based assistive technology to determine whether the defined gap exists. We followed the steps shown in Kitchenham et al. systematic literature review~\cite{kitchenhamsystematic2009} for this work. 

\subsection{Research Questions}

The aim of conducting this systematic review is to address the following research question:

How are failures and risks explored, assessed, and reported in research on AI-based assistive technologies? \\Eight more detailed questions are derived from this research question, expressed in the following sections. Together, these cover various aspects of our research question.
\subsection{Questions}
\begin{figure}[h]
    \centering
    \includegraphics[width=0.5\textwidth]{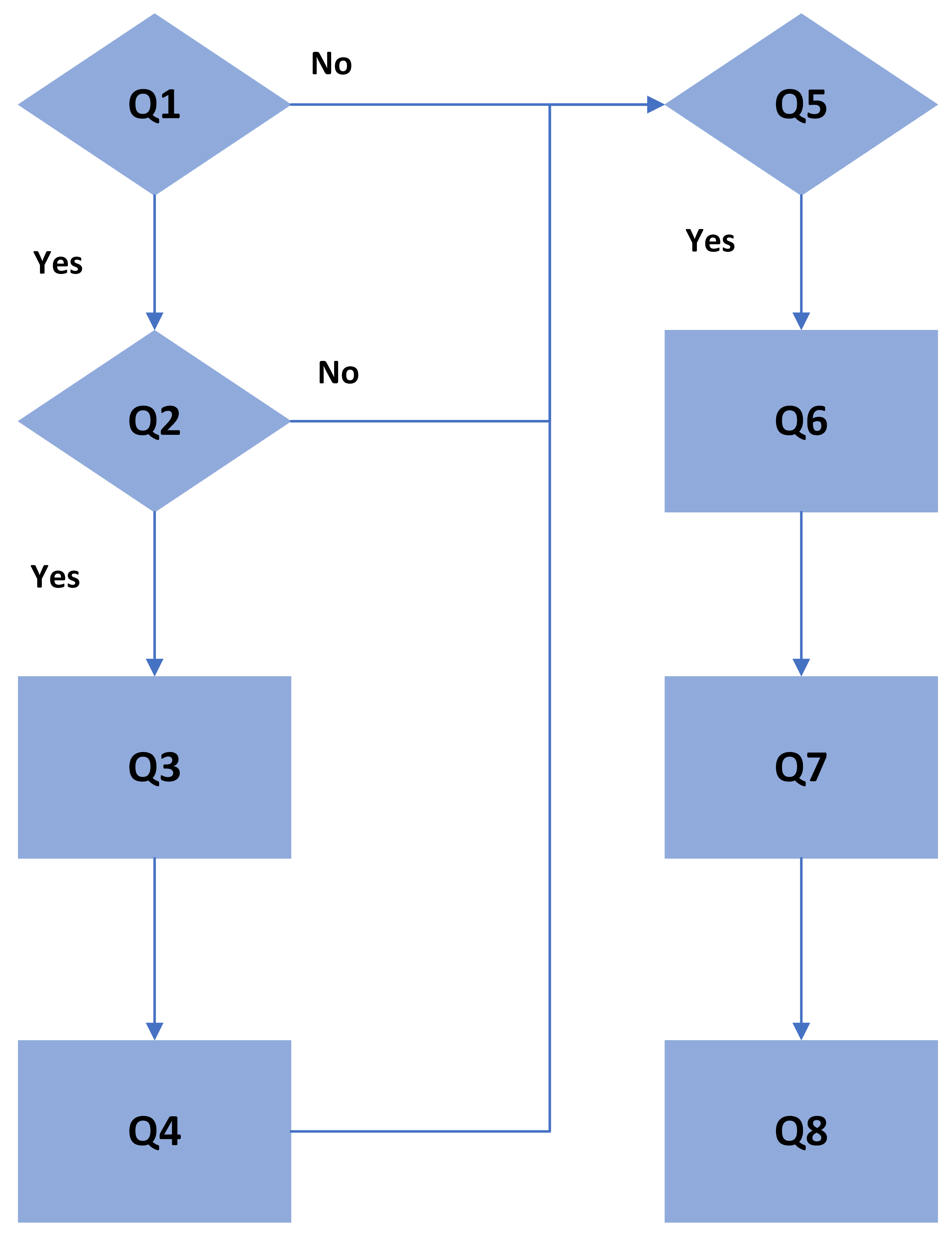}
    \caption{
    Question flow chart - Questions 1, 2, and 5 are represented as conditional nodes. If Question 1 is answered 'yes,' the flow continues to Question 2; if 'no,' it jumps directly to Question 5. Similarly, if Question 2 is answered 'yes,' the flow continues to Questions 3 and 4 before moving to Question 5; if 'no,' it also jumps directly to Question 5. If Question 5 is answered 'yes,' the flow proceeds to Questions 6, 7, and 8. Each decision point is guided by a 'yes' or 'no' response, leading to a structured progression through the research questions.
    }
    \label{fig:questionChart}
    
\end{figure}

To analyze the papers, we reviewed each paper by answering eight assessment questions derived from the main research question above. These are two parallel sets of questions as shown in Figure\ref{fig:questionChart}. The first four questions focus on the human study and the contribution of the sight-loss community to the study, and the following four questions are about failure considerations of the represented system. By the word "system," we mean any tool, device, or method that is presented in the paper. The assessment questions are as follows: 

\begin{enumerate} 
\item Does the paper present a working demo or examples of how the system behaves?
\item Is there an evaluation of the system by conducting a human study?
\item Is the human study ecologically valid for persons with visual impairment? In other words, does the human study include members of the sight-loss community? 
\item Does the paper consider any threats to the validity of the human study?
\item Does the paper provide evidence that the authors considered risks/possible failures associated with the system?
\item Does the paper report examples of failures?
\item Does the paper give specific information about when and how the system will fail?
\item Does the paper talk about the consequences of the reported/systems failures?
\end{enumerate}

\subsection{Definitions and Scope}

To proceed with the work, the definition of "failure" in assistive technologies is specified in this section because there might be confusion about a work's failure, limitation, or disadvantage while reading and analyzing papers. To see what can be considered a failure, the following definitions are helpful:

The definition of the word "failure" in Oxford English Dictionary~\cite{OxfordDictionary} is explained as "lack of success", "breaking down or ceasing to function". In engineering, a system's failure could refer to a component's inability to operate as expected~\cite{brooksfailure2002}. Regarding AI systems, an AI incident is defined as "a situation where AI systems caused, or very nearly caused, real-world harm"~\cite{mcgregorpreventing2021}. Williams and Yampolskiy~\cite{williamsunderstanding2021}, defined AI failure as the "Malfunctioning of an AI system". We distinguished between failure, limitation, and disadvantage of the system following three simple definitions:

\textsl{Failure}: The system is intended to do x but does y instead or does not do anything at all.

\textsl{Disadvantage}: The system does x, but there are undesirable things about doing x.

\textsl{Limitation}: The system is intended to do x and y but only does x.\\

While we report the results of analyzing each paper in the following sections, only failure is in the scope of this review, not disadvantage and limitation. It also would be valuable to do a review on the limitations and disadvantages of the systems in the future.
In this study, our focus is on AI-based assistive technologies, so to distinguish between AI-based systems and other assistive software and technologies, we refer to Canada's Artificial Intelligence and Data Act (AIDA) definition of AI~\cite{bill}:

\textsl{"artificial intelligence system means a technological system that, autonomously or partly autonomously, processes data related to human activities through the use of a genetic algorithm, a neural network, machine learning or another technique in order to generate content or make decisions, recommendations or predictions."}

\section{Related Work}
A few Systematic Literature Reviews have been done in the field of assistive technologies. For instance, Alper and Raharinirina \cite{alperassistive2006} discussed that although assistive technology devices are becoming more common, we do not know enough about how they are used by individuals with different types and levels of disabilities and at different ages. To address this issue, the authors reviewed published reports on assistive technology for persons with disabilities and analyzed the uses, benefits, and obstacles of such technology. The study results provide insight into why technology is often used less than it could be.

Boot et al. \cite{bootaccess2018}, conducted a systematic literature search to identify barriers and difficulties in utilizing assistive technology for individuals with intellectual disabilities. The most frequently reported barriers were lack of funding, needing to know more about the technology, and needing to be appropriately assessed.

Tyagi et al.~\cite{tyagiassistive2021}, have done a review on navigation systems designed for visually impaired and blind persons. They have gathered papers from the years 2001 to 2020. Although all these papers have introduced different assistive technologies for visually impaired persons' navigation, they still have their limitations. These limitations can be due to the performance issue of the devices under certain circumstances. Such as using GPS only works in outdoor scenarios and WIFI-based approaches only for indoor scenarios. Some of the proposed assistive technologies' accuracy is not as precise as it should be, and for some others, the high cost of the device is the issue. Finally, the paper has concluded that the device's result depends on humanitarian conditions and the task of designing an assistive technology for navigation purposes in real life is challenging.

Khan et al.~\cite{khananalysis2021}, conducted a systematic review of the assistive technologies proposed for visually impaired persons. They have reviewed 191 different articles from the years 2011 to 2020 from 6 different databases. Their research questions are mostly related to the proposed navigation algorithm, the technologies used, and the mechanisms for avoiding obstacles in the path, and the method each article has validated in terms of its applicability and reliability. The results for the final research question prove most of the articles suffer from a lack of a reliable validation approach.\\

In the field of AI safety and failures, Williams and Yampolskiy~\cite{williamsunderstanding2021} developed a framework applicable to AI systems to assess their safety level when used in practical scenarios. The framework examines various aspects of the system, such as observability and correctability scores. Using the outcome of this examination, the authors defined some rules followed with recommendations to enhance the system's overall safety.

Amodei et al.~\cite{amodeiconcrete2016} have focused on unintended and harmful behaviours AI systems can bring to this world. Although they have mostly focused on applications related to reinforcement learning, part of the study can be generalized for any AI-driven system. Such as how to detect whether our system is disturbing the environment, the failure costs, creating life-threatening situations, or the system robustness in different environments and facing different data distributions. Other safety-related issues such as privacy, fairness, security, and transparency are also discussed. The study concludes that with the recent rapid growth of AI there are more systems controlling industrial processes,
health-related systems, and other mission-critical technology. Therefore, small incidents are very concrete threats that need to be prevented. Otherwise, despite the immediate harm, it can also cause justified loss of trust in automated systems.\\

Although there are a considerable number of papers proposing assistive technologies for persons with disabilities and the AI-driven approach failures and threats are concerning, there is a gap in the literature to review failure threats and risks of AI-based assistive technologies for persons with visual impairment. Therefore, in this paper, we have evaluated this topic.

\section{Methodology}
\subsection{Systematic Search and Data Extraction}
The details of the search process for finding the target papers are provided in this section to clarify the steps of the data collection process, avoid biased assumptions, and tackle the intention to choose papers that appeal to the expected result.
The search process included a manual search for papers through Scopus, IEEE Xplore, and ACM digital library databases using related keywords and search strings. It is important to choose a search string carefully since a word with various meanings can make a noticeable difference in the results. Furthermore, we added "AND NOT" terms to our search string to to avoid capturing papers related to diagnosis and treatment. The final search string is as follows:

\textsl{( ( assist*  OR  guide  OR  tool  OR  technology )  AND  ( "artificial intelligence"  OR  ai  OR  "machine learning"  OR  smart )  AND  ( "visually impaired"  OR  "blind people"  OR  "blind person"  OR  "sight loss"  OR  "vision loss"  OR  "partially sighted" )  AND NOT  ( screening  OR  diagnos*  OR  treatment  OR  medicine  OR  epidemiology ) )}

\subsection{Inclusion/Exclusion and Sampling}
After finalizing the search string, we used it in Scopus, ACM digital library, and IEEE Xplore advanced search. We also explored using Google Scholar initially but found the search functionality to be less precise than the other databases. Therefore, we excluded Google Scholar from our data search since advanced searching and exporting features are more comprehensive and diverse in the first three databases. We searched in three categories of: paper titles, abstracts, and keywords. Moreover, we filtered all the results for the past five years (2017 - 2022). We excluded 2023 since it was the ongoing year when we started searching for the papers, and the number of published papers was still increasing. We excluded years before 2017 since we are focusing on AI-based technologies, and in recent years, there has been a rapid growth of new AI systems that are yet to be understood entirely.

After collecting the results, there was still the chance of having irrelevant papers, so we reviewed all the paper titles, in some cases abstracts, to classify all the results as "Relevant," "Irrelevant," and "Related" papers. "Related" papers are studies engaging in the topic but do not present an assistive technology. These papers include surveys, reviews, or proposals related to the subject. "Relevant" papers as our final dataset for the systematic analysis include papers that developed or improved an AI-based assistive technology for persons with visual impairment. 

\subsubsection{Sample papers sets}
Since reviewing all 648 papers proved infeasible, in order to ensure both the feasibility and validity of the review, we adopted a sampling approach. 
We initially started our review with randomly chosen sample papers in our "Relevant" class in each of Scopus, ACM, and IEEE Xplore results. As we continued reviewing the papers, we found that the results for each question were consistent, and percentages remained nearly the same, suggesting that increasing the sample size would not add a notable difference to the results. Considering this, we decided to preserve the sampling approach and show the findings of a subset of 100 papers. We separated these 100 papers into two sets of 50 to ensure the results were similar in the two sets so we could conclude if our samples were representative of the total papers. While choosing these sample sets, we eliminated some more papers after reading them because they were outside our scope of work; either they were not AI-based or did not represent a new assistive technology.
\subsubsection{Sample sets validity}
\begin{figure}[h]
    \centering
    \includegraphics[width=0.7\textwidth]{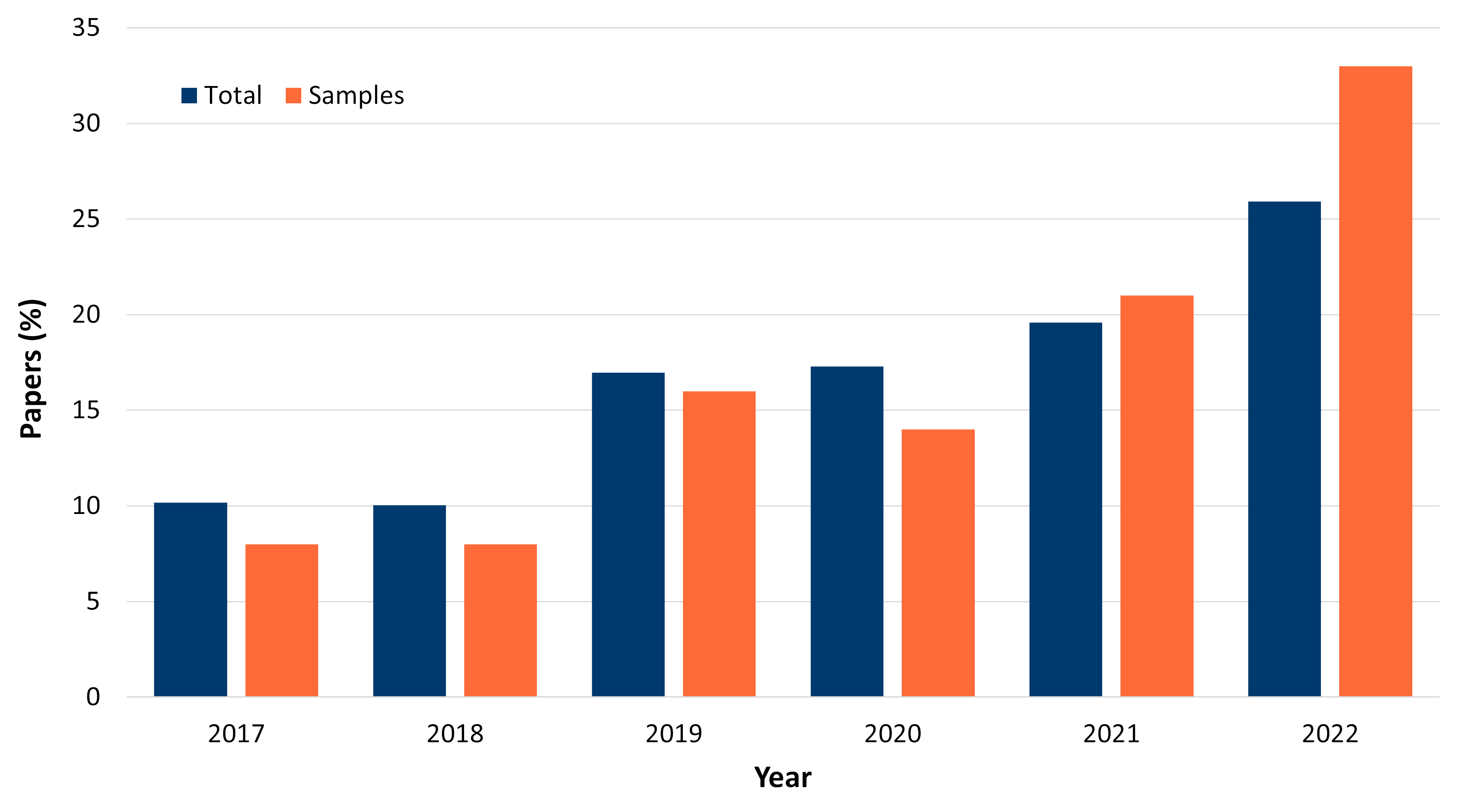}
    \caption{
    Bar chart showing the yearly distribution of total papers and sample papers, spanning from 2017 to 2022. The x-axis represents the years in descending order from 2022 to 2017, while the y-axis indicates the percentage of papers. The bars compare the percentage of papers in the sample set to the total set for each year. The closeness of percentages between the sample and total sets for each year shows our sample papers are representative of the total number of papers. Also, it shows that the number of papers has increased over the years.
    }
    \label{fig:year}
\end{figure}

To ensure that our sample papers are representative of the total number of papers, some attributes of the papers in sample sets and the total papers have been compared. We chose the papers randomly with a uniform distribution. Furthermore, as it is shown in the Figure~\ref{fig:year} the sample closely tracks the distribution of publication years in the full set.

\subsubsection{Questions validation}

To validate the methodology and the efficacy of the questions, we examined how straightforward it was to assess each paper, to ensure each designed question is answerable.\\

\textsl{Q1. Does the paper present a working demo or examples of how the system behaves?}\\

Addressing this question was straightforward and with no complexity. It is worth mentioning that in "How the system behaves," we did not mean the methodology or details of the system's architecture, but we meant the examples of using the system and produced outputs.\\ 

\textsl{Q2. Is there an evaluation of the system by conducting a human study?}\\

For this question, no matter how many persons are involved, if the paper represents a user study, we consider it as a positive answer. Addressing this question did not pose any challenge since, in each study, if there is a human study, they had mentioned it clearly.\\ 

\textsl{Q3. Is the human study ecologically valid for persons with visual impairment?}\\

In some of the studies we checked, there was a human evaluation, but it was done with blindfolded participants, not members of the sight-loss community. This evaluation method is inappropriate since the situation is temporary and is not psychologically accurate for the participant. Specifically, it does not fully encapsulate the long-term or psychological effects associated with vision loss. Moreover, it does not account for a variety of vision loss levels. Instead, it presupposes a certain type and quality of vision loss in the use of a blindfold. Such an approach is, in fact, a simulation that can lead to invalid conclusions and be discriminatory~\cite{silvermanperils2015}. Determining whether the users in the human study are blindfolded or persons with visual impairment was relatively easy since the authors clearly mentioned this. However, answering this question positively could not be only "yes" because the number of participants and the level of sight can vary, such that the human study can represent visually impaired users at different levels. It could be a total population of participants with visual impairment or only one blind person testing the assistive tool. So, the question could be asked: How is human study ecologically valid for visually impaired persons? It can be answered as "It is not ecologically valid," "Partially," and "Completely." However, the challenge is determining what constitutes partial or complete validity in a given application. So, we answered positively if there was a study that involved the sight-loss community in some way, and this is an intentionally low bar. A further study exploring the validity of each human study would prove valuable in the future.\\

\textsl{Q4. Does the paper consider any threats to the validity of the study?}\\

What makes it challenging to answer this question is that in the selected papers, we did not observe a separate section or analysis for threats to the validity of work; however, this validation might be indirectly presented in other sections so it is needed to analyze the whole paper and extract the mentioned threats to the validity of work, or be content with what we get with just having a look at the paper.\\

\textsl{Q5. Does the paper provide evidence that the authors considered risks/possible failures associated with the system?}\\

This question has been addressed in different ways in the studies, such as providing examples of failures or sometimes by explaining how the systems might fail. 
We also answered this question positively if the paper reported accuracy, which means the author/s are aware that there is a chance that their system will fail.\\

\textsl{Q6. Does the paper report examples of failures?}\\

This question was answered yes if the paper provides examples of the system's failure while testing the system or conducting the human study. \\

\textsl{Q7. Does the paper give specific information about how and when the system will fail?}\\

Answering this question depends on whether the paper talks about possible situations in which the system might fail or how it would be if it fails. 
So, in some studies, they considered that their system might fail (answered yes to Question 5) by providing some examples of failure (i.e. Question 6) or calculations of the accuracy. However, these questions are not concerned with identifying situations when the system fails; this is the focus of Question 7.
Some papers answered Question 7 based on their answer to Question 6. Here, the paper reported examples of system failures and based on those reported failures, talked about the situations in which they can happen. Other papers did not report examples of failures (answered No to Question 6). However, they explained the possible situations in which the system might fail. These could be theoretical expectations or based on unreported failures.
As an instance of positively answering this question, many papers that presented an image recognition system suggested that the system might not generate accurate results if the picture quality is low or blurry. So, if image detection is used in a navigation app in foggy weather, it may fail.

\textsl{Q8. Does the paper talk about the consequences of the reported/systems failures?}\\

This question was sometimes challenging to answer because, usually, there was no direct referencing to the consequences of a system failure, but the authors talked about it indirectly. Interpreting this from a failure report requires personal judgment.
\section{Results and Analysis}
\subsection{Systematic search results}

After cleaning up our search results, as noted in the inclusions and exclusions section and removing the overlapped papers from three databases, 648 studies were collected as studies that present an assistive technology for persons with visual impairment.
While doing the systematic review, we eliminated more papers since, by analyzing each study, we discovered more details about the paper meeting the inclusion/exclusion criteria, like if the system is AI-based and compatible with the reference definition of AI.
In our initial filtering stage, we needed more certainty regarding the status of many papers. Therefore, it remains necessary to perform additional filtering as part of a complete systematic review in the future.

\begin{table}[]
\centering
\begin{tabular}{|c|c|c|c|c|}
\hline
\textbf{Source} & \textbf{Retrieved after Inc. Exc.} & \textbf{Manual filter} & \textbf{Unique} & \textbf{Samples} \\ \hline
\textbf{ACM}    & 38                        & 17                     &                 &                  \\ \cline{1-3}
\textbf{IEEE}   & 296                                & 254                    & 648             & 100              \\ \cline{1-3}
\textbf{Scopus} & 970                                & 603                    &                 &                  \\ \cline{1-3}
\textbf{Total}  & 1304                               & 874                    &                 &                  \\ \hline
\end{tabular}
\vspace{0.5em}
\caption{Number of papers retrieved from different sources and number of sample papers}
\label{tab:papers}
\end{table}

\subsection{Systematic review results}

The results of applying the assessment questions for each of the paper sets are presented in Table~\ref{tab:commands}. As shown in the table, the results for both sets are close and do not have significant differences, which shows that these sample papers are representative of the whole paper. It also indicates that the mentioned gap about the lack of study on risks and failures associated with assistive technologies exists, and this systematic review demonstrated this. In the following subsections, the results for the total number of a hundred papers are discussed separately for each question.

\begin{table}[h]
  \centering
  \begin{tabular}{c|cccccccc}
  \hline
     & Q 1 & Q 2 & Q 3 & Q 4 & Q 5 & Q 6 & Q 7 & Q 8\\ \hline
    \addlinespace
    \textbf{Set 1} & 86\% & 28\% & 16\% & 6\% & 70\% & 24\% & 32\% & 2\%\\
    \addlinespace
    \textbf{Set 2} & 84\% & 26\% & 16\% & 2\% & 78\% & 20\% & 32\% & 2\%\\ 
    \addlinespace
    \textbf{Total} & 85\% & 27\% & 16\% & 4\% & 74\% & 22\% & 32\% & 2\%\\ \hline
  \end{tabular}
  \vspace{0.5em}
  \caption{Preliminary findings from the systematic analysis of two sets of sample papers: The number for each question shows the percentage of how many studies among our sample papers address that question in their work}
  \label{tab:commands}
\end{table}

\begin{figure}[H]
    \centering
    \includegraphics[width=0.7\textwidth]{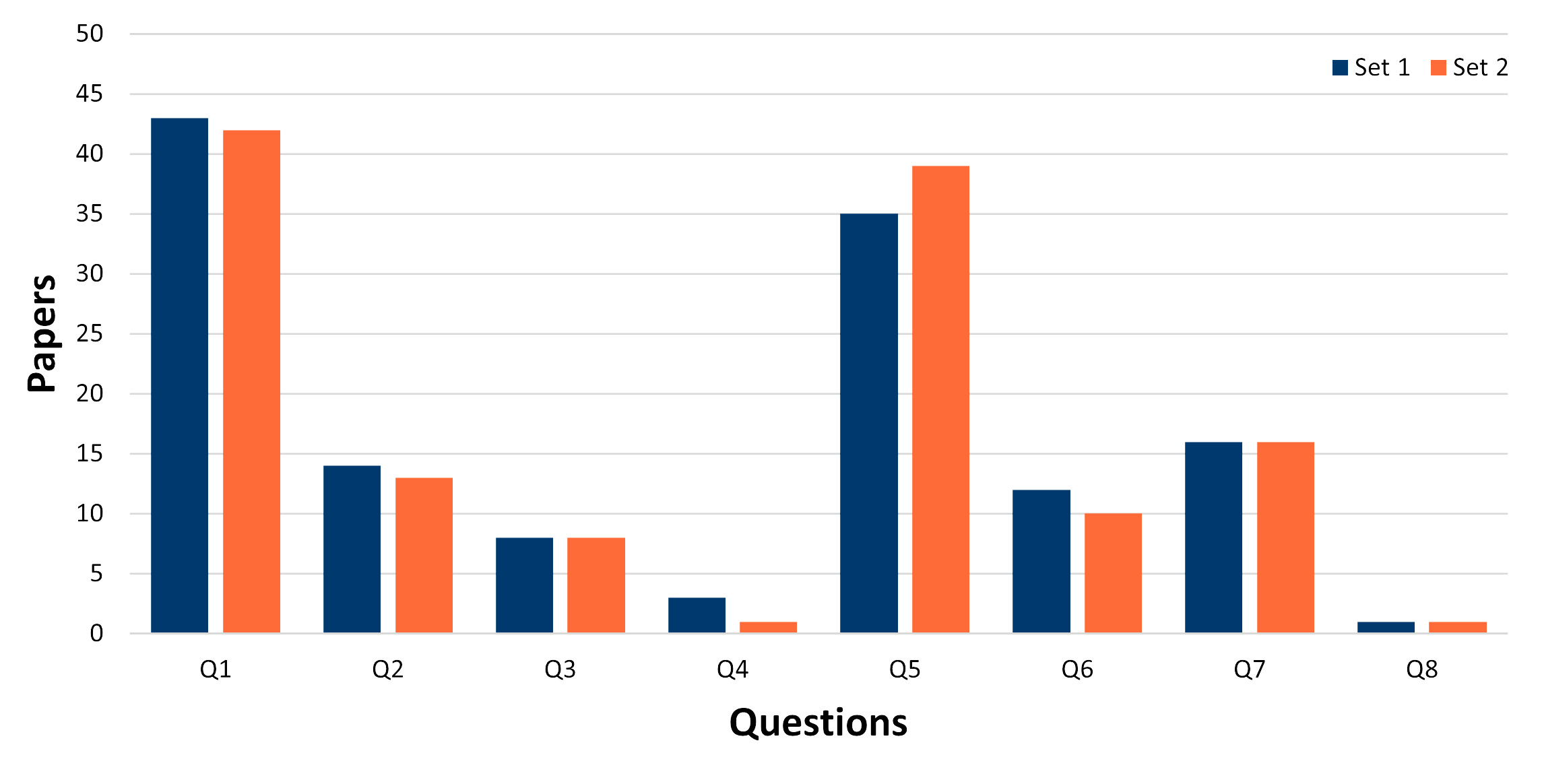}
    \caption{
    Final results for two sets of papers
    }
    \label{fig:question results}
\end{figure}

\subsubsection{Question 1: \textsl{Does the paper present a working demo or examples of how the system behaves?}}
\begin{figure}[H]
    \centering
    \includegraphics[width=0.5\textwidth]{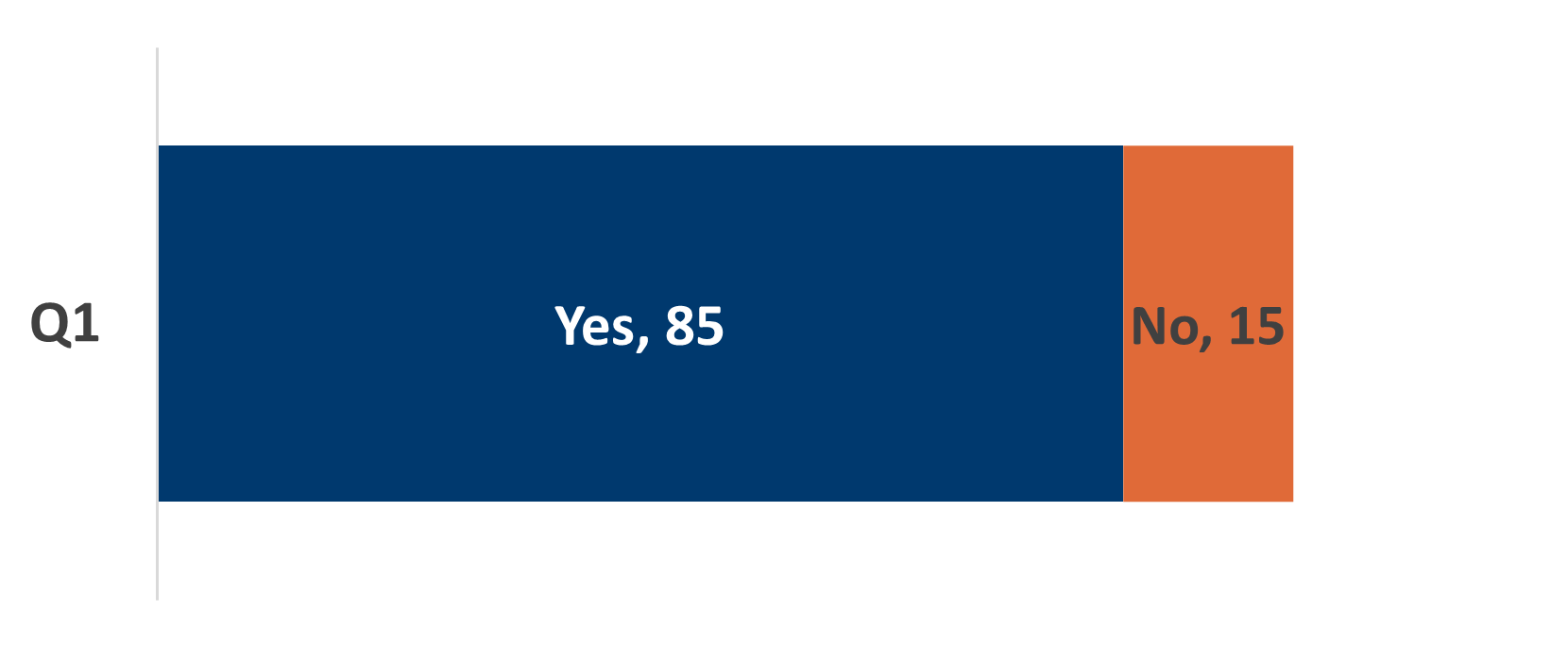}
    \caption{
    The number of papers in the combined set according to their answer to Question 1: 85 Yes, and 15 No
    }
    \label{fig:Q1}
\end{figure}

As shown in figure~\ref{fig:Q1}, 85 papers have answered positively Question 1, meaning they have a working demo or an output of the system's behaviour. This could be in the form of a picture of the system's output or showing the specific results of a test case. The other 15 papers did not address this question, whether they do not represent any output or the question does not apply to them, for example, in cases of studies that are in the early steps or are proposals.
\subsubsection{Question 2: \textsl{Is there an evaluation of the system by conducting a human study?}}
\begin{figure}[H]
    \centering
    \includegraphics[width=0.5\textwidth]{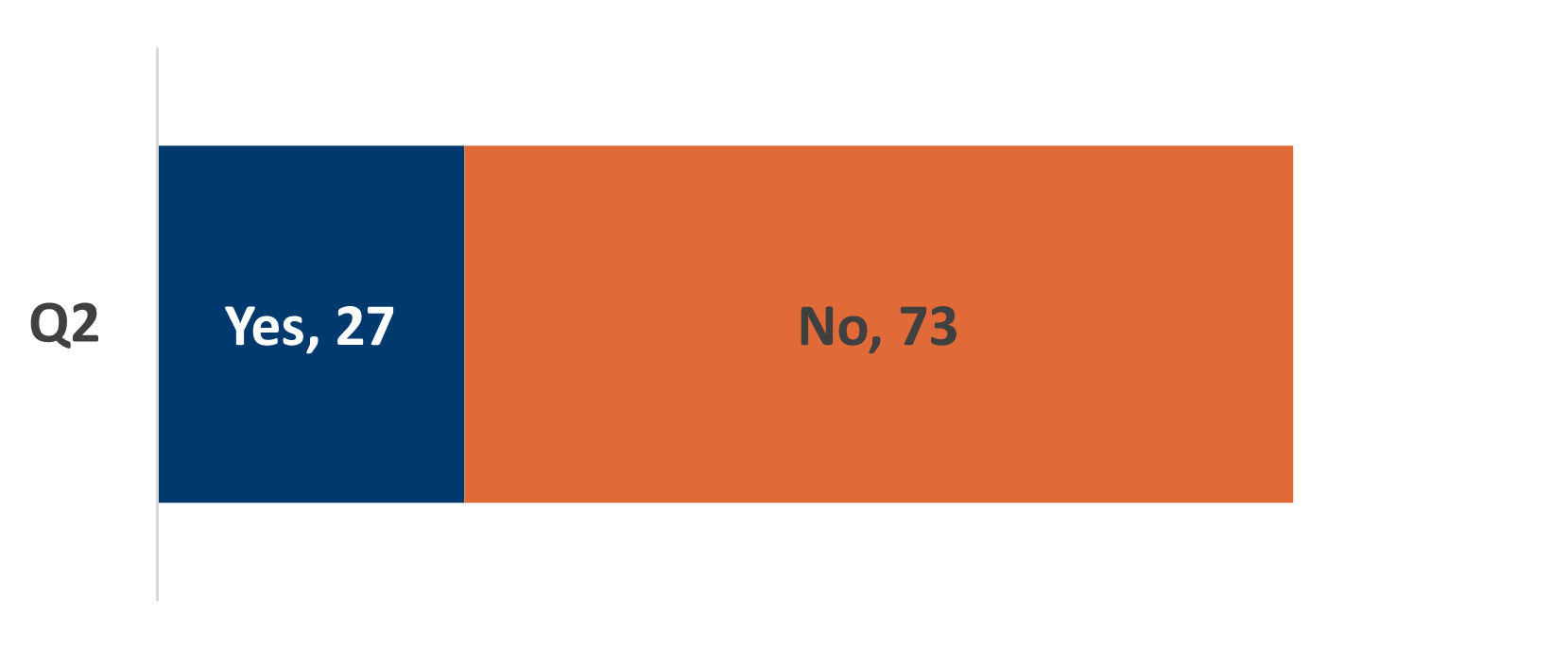}
    \caption{
    The number of papers in the combined set according to their answer to Question 2: 27 Yes, and 73 No
    }
    \label{fig:Q2}
\end{figure}

As demonstrated in figure~\ref{fig:Q2}, among the papers with an actual product or prototype to be tested, only 27/85 = 32\% of them have been evaluated by human experiments. This result suggests that a noticeable part of innovations are only evaluated based on the performance and theoretical outputs. These technologies are intended to assist people. Without human trial, it is not possible to show if the system is actually useful for real-life tasks, so it does not serve its intended purpose of assisting people. Moreover, user-friendliness and the way each person interacts with it play an important role in the result. The system output might be correct, but the user is still unable to understand the result or make any decision.
\subsubsection{Question 3: \textsl{Is the human study ecologically valid for persons with visual impairment?}}
\begin{figure}[H]
    \centering
    \includegraphics[width=0.5\textwidth]{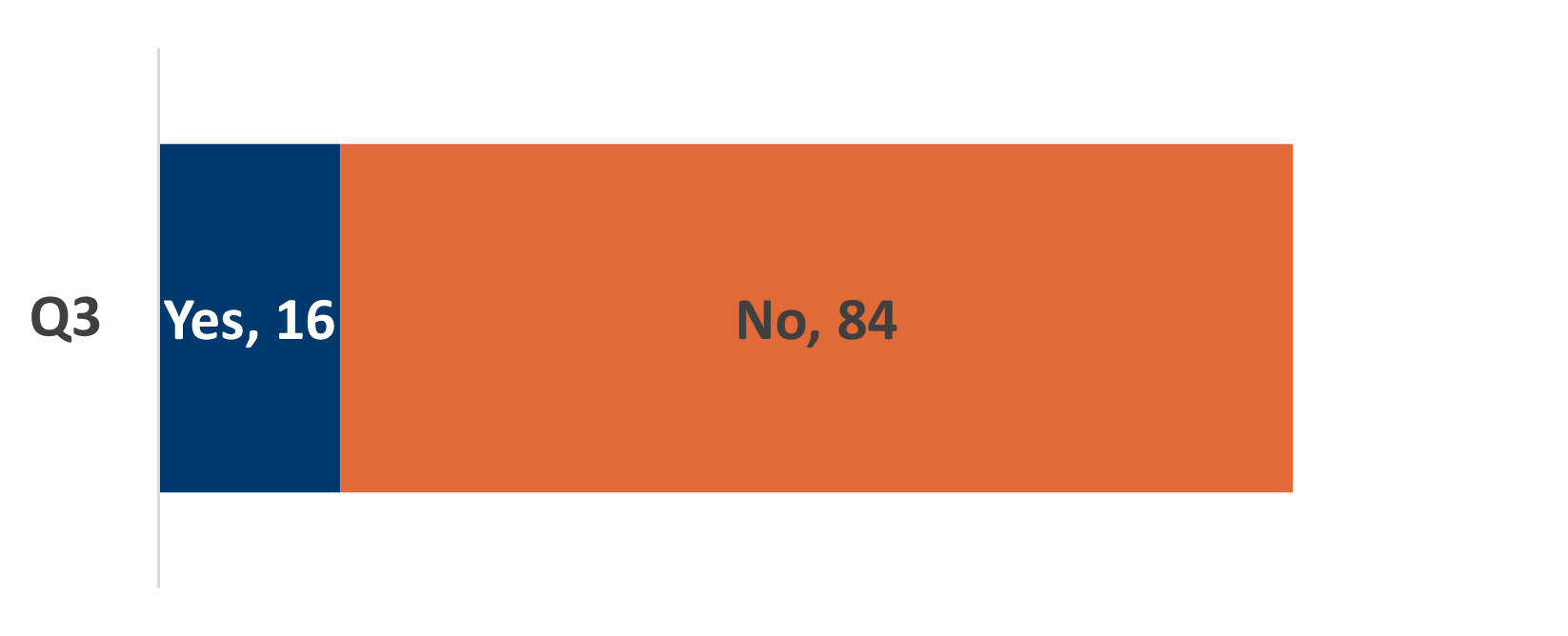}
    \caption{
    The number of papers in the combined set according to their answer to Question 3: 16 Yes, and 84 No
    }
    \label{fig:Q3}
\end{figure}

As shown in figure~\ref{fig:Q3}, among the papers that answered yes to Question 1, only  16/85 = 19\% of them have had at least one participant with visual impairment for the experiments. This suggests that, in total, out of all the papers that were checked, only a few of them have been evaluated by persons from the sight-loss community who are the target users. So, the theoretical evaluation of the system does not represent the community targeted by the research. Also, compared to the answers to Question 2, five papers had human study, but none of the participants were from the sight-loss community, and sighted or blindfolded participants did the test.
For example, Akkapusit and Ko~\cite{akkapusittaskoriented2021} have conducted a human study with 20 participants of different ages, genders, and experience levels using the provided equipment for the evaluation. However, all participants were sighted and blindfolded for the study.
This is not an appropriate approach to evaluate an assistive technology intended to be used by the sight-loss community. 
\subsubsection{Question 4: \textsl{Does the paper consider any threats to the validity of the human study?}}
\begin{figure}[H]
    \centering
    \includegraphics[width=0.5\textwidth]{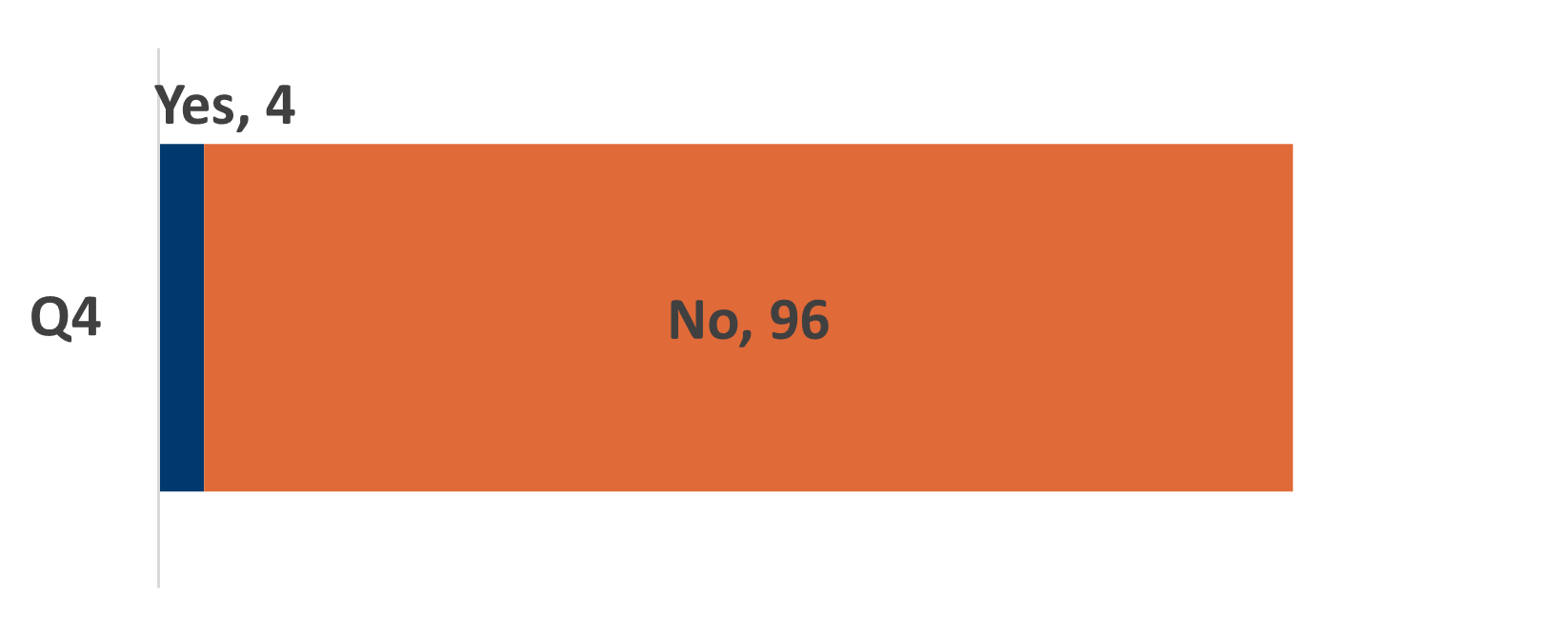}
    \caption{
    The number of papers in the combined set according to their answer to Question 4: 4 Yes, and 96 No
    }
    \label{fig:Q4}
\end{figure}

Another important aspect that needs to be considered while testing is the validation of the result. Figure~\ref{fig:Q4}, shows that only 4\% of papers have considered any threats to the validity of their human evaluation. The system evaluation depends on the different criteria of human study, such as participants' age ranges, heights, education levels, specific weather conditions, light conditions, and many more criteria. As a result, these evaluations can not be fully representative of their target users, and only a few research studies have considered this fact. 
As an instance of considering threats to the validity of the work, Easley and Rahman~\cite{easleydeveloping2021} mentioned that their study could be improved by recruiting participants with a diversity of attributes like vision level, age, gender, experience, etc.
\subsubsection{Question 5: \textsl{Does the paper provide evidence that the authors considered risks/possible failures associated with the system?}}
\begin{figure}[H]
    \centering
    \includegraphics[width=0.5\textwidth]{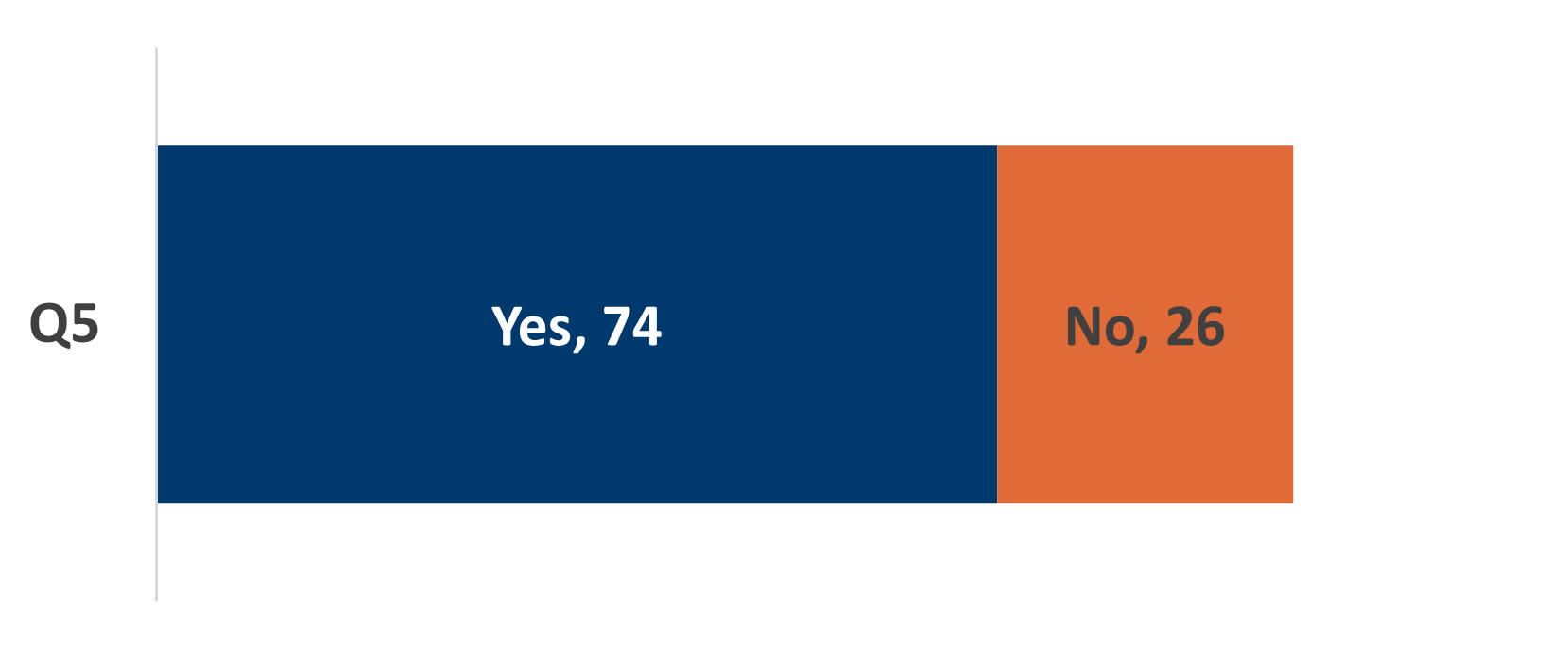}
    \caption{
    The number of papers in the combined set according to their answer to Question 5: 74 Yes, and 26 No
    }
    \label{fig:Q5}
\end{figure}

In figure~\ref{fig:Q5}, we have examined whether the papers have considered the probability of failure for their proposed system. As shown, 74\% of papers have at least introduced a metric to evaluate their proposed approach or prototype success rate. It is important to mention that this performance report does not necessarily rely on human experiments. It can be the system behaviour from different points of view, such as time efficiency, accuracy and the correctness of the results. We considered this question addressed by the paper, even if it only reports the system's accuracy, thereby acknowledging that failure is a possibility. 

\subsubsection{Question 6: \textsl{Does the paper report examples of failures?}}
\begin{figure}[H]
    \centering
    \includegraphics[width=0.5\textwidth]{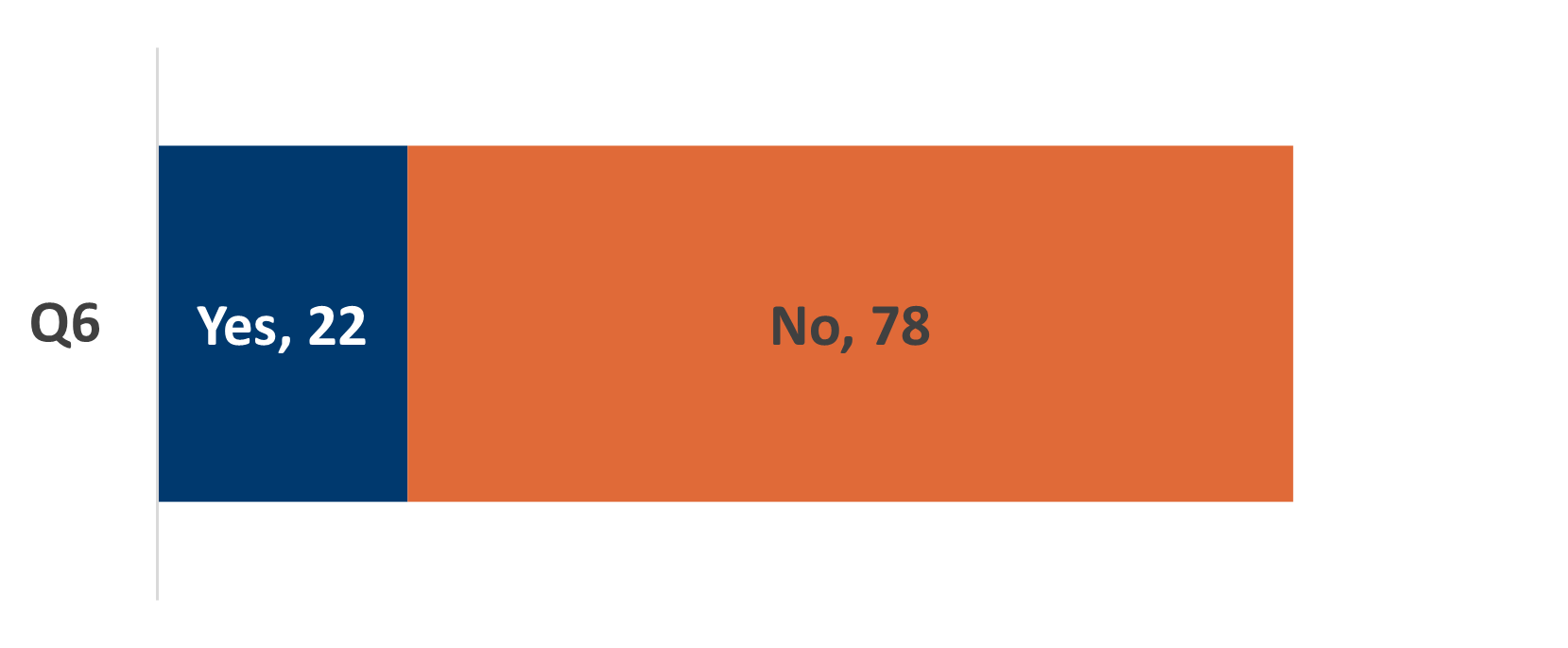}
    \caption{
    The number of papers in the combined set according to their answer to Question 6: 22 Yes, and 78 No
    }
    \label{fig:Q6}
\end{figure}

As the results demonstrate in figure~\ref{fig:Q6}, 78\% of the papers did not give an example of their system failure and incorrect results. To use these systems daily, we need to know about the cases in which they have failed. Ignoring the failure samples will just be like ignoring an existing issue that will never allow the proposed system or approach to be used in real-life scenarios. 

\subsubsection{Question 7: \textsl{Does the paper give specific information about when and how the system will fail?}}
\begin{figure}[H]
    \centering
    \includegraphics[width=0.5\textwidth]{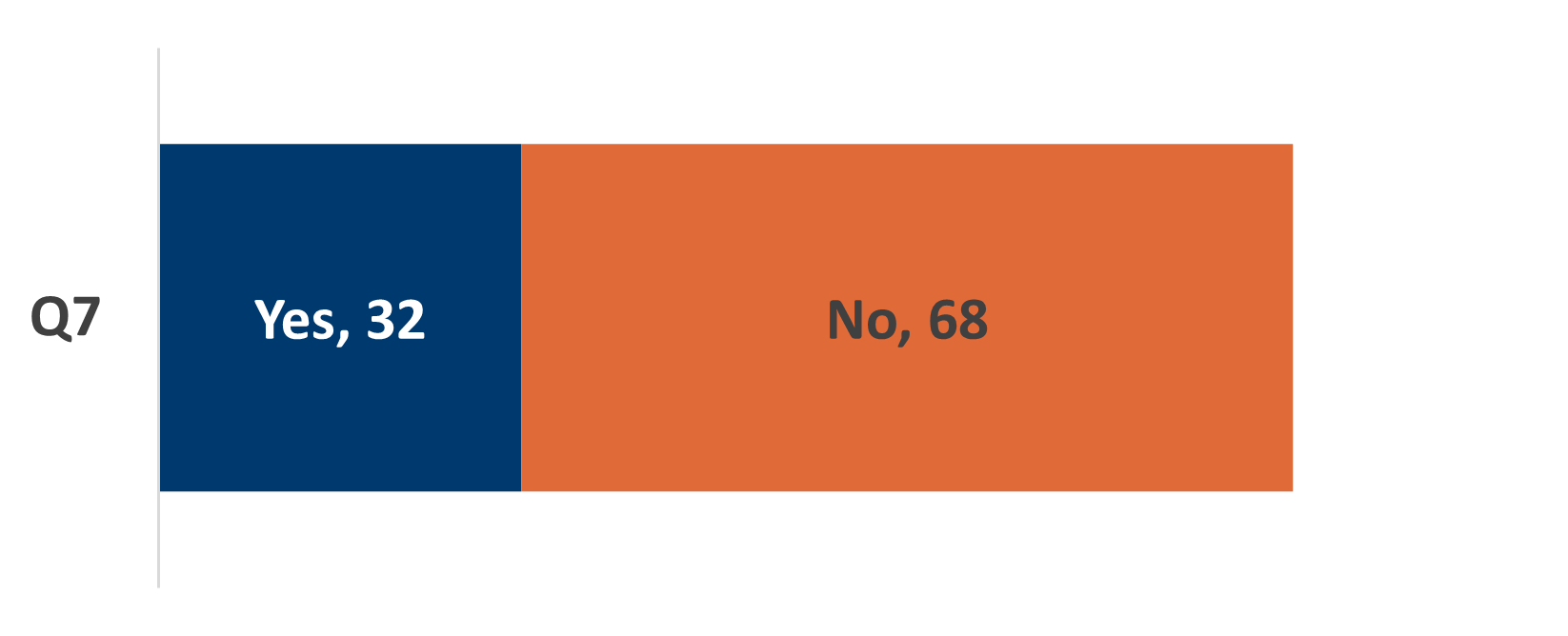}
    \caption{
    The number of papers in the combined set according to their answer to Question 7: 32 Yes, and 68 No
    }
    \label{fig:Q7}
\end{figure}

As figure~\ref{fig:Q7} shows, 32\% of papers have explained the situations in which their system will malfunction and fail. This can be a certain brightness level, background noises, users' special conditions, and many more environment or user-related features. A system can be robust and reliable within a certain setup and not the other. For example, a navigation system may function perfectly outdoors and not indoors. Knowing the situation that the system might fail can help the user decide when to rely on it and when not to.
\subsubsection{Question 8: \textsl{Does the paper talk about the consequences of the reported/systems failures?}}
\begin{figure}[H]
    \centering
    \includegraphics[width=0.5\textwidth]{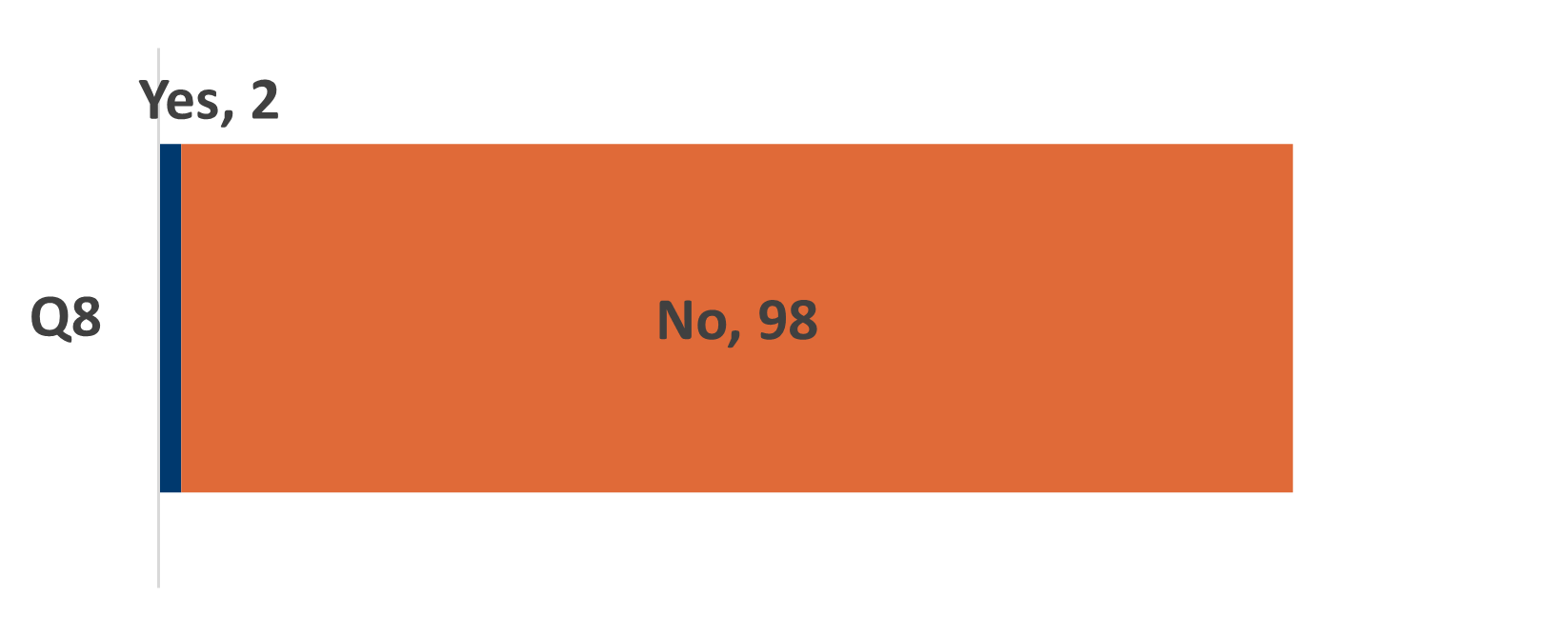}
    \caption{
    The number of papers in the combined set according to their answer to Question 8: 2 Yes, and 98 No
    }
    \label{fig:Q8}
\end{figure}

In figure~\ref{fig:Q8} it has been shown that only 2\% of papers have considered their system failure consequences. Referring to Question 5, it shows that most studies acknowledge and are aware that their system might fail, but very few of them talk about the existing or potential consequences of these failures. 
An assistive system to help persons with visual impairment can directly affect one's well-being. So, it needs to be robust and have some error-handling mechanisms when they fail. This has only been considered in a few papers. For any system to be used in real-life situations, the consequences of failure must have been considered and handled to get close to being used in reality despite any performance or evaluation challenges. This handling could mean trying to mitigate the risks, warning users not to use it in certain situations, or letting users decide whether they want to trust the assistive technology with the new information. 
As an example of this question being positively answered, Li et al.~\cite{araiCrossSafe2020} discussed that the detection system needs a high accuracy since it has to detect crucial situations like a red traffic light. 

\section{Discussion}
Most studies demonstrated a working example of their system's behaviour; however, among these papers, only a few evaluated their working on an actual test with human participants. Even among those who have a human experiment, some papers had blindfolded participants instead of persons with visual impairment. This and many other factors mentioned in 4.2.4 question the validity of human studies, but very few papers talk about threats to the validity of their human study.

On the failure consideration side, most studies acknowledged that their system could create incorrect outputs (Question 5). However, again, only a tiny percentage of these papers specified the situations in which the system might fail or gave examples of the system's incorrect results.
Moreover, By analyzing the review results, we noticed that among the papers without human studies, only 13.7\% of them reported the system's failure cases. In comparison, among the papers with human studies, 44.4\% reported examples of their system's failure. This number is 50\% for the studies that had a human study with participants from the sight-loss community. This comparison suggests that without the presence of a human study, the chance of having unnoticed failure cases increases.
In the end, only a few papers give awareness about what might happen due to the system's failure to clear the associated risks of using the product. 

As our systematic review shows, there is a need to revisit how the academic research on assistive technologies equipped with AI is done and presented. Now that we know this need exists, it is necessary to determine how this kind of research should be done and what steps should be followed to help make sure studies are valid and represented in a way that allows the user to decide about using assistive technology and trusting it. Designing guidelines to follow is our potential next step. We know that the gap in this research area exists, and now we look forward to a solution that will help minimize the gap. 

\section{Limitations}

A limitation of this work is narrowing down our context to academic studies and published papers. We acknowledge that there are assistive technologies invented or provided by companies that do not publish in academic venues. Still, as stated in our research question, we were primarily interested here in exploring academia and research on AI-based assistive technologies. Furthermore, the final search string was arrived at in order to be comprehensive while avoiding excessive irrelevant results. Still, no perfect search string could cover all the related papers. So, it is possible that some papers are not captured with our search string and are not included in our search results.

A hundred papers among the total 648 papers were examined as a sample of papers. Although they were chosen randomly and showed that this sample is representative of the whole, it is still expected that the results would vary slightly if all 648 papers were assessed. 
Also, we did the assessment manually, and there is always the possibility of introducing subjectivity and unconscious bias. 
\section{Conclusion}
In this work, we conducted a systematic literature analysis of the published studies to understand how the failures and risks of AI-based assistive technologies for persons with visual impairment are presented in academia. 
We framed eight research questions, which were then applied to a subset of 100 papers randomly sampled from the total number of 648 papers in the systematic search results. The questions captured both human study and risk analysis aspects of the study. Results show only a few papers (19\% among those that have a prototype or demo of the system) tested their work in a human study with visually impaired participants. Moreover, there is less information on the cases of failures and possible risks of introduced systems. We validated the method as the search string and inclusion/exclusion criteria effectively captured a considerable amount of relevant literature within our research scope, and the proposed analysis questions were answerable and provided meaningful results. 

In our future work, we look to take the findings of this research to introduce guidelines for researchers on how to evaluate AI-based assistive technology well. This will ultimately empower people to judge the trustworthiness of a newly presented assistive technology.

\section{Acknowledgements}
We would like to show our appreciation to members of the Canadian National Institute for the Blind (CNIB) for the valuable conversations and insights that brought us a better understanding of accessibility challenges and barriers. This research was undertaken, in part, thanks to funding from the Canada Research Chairs Program. This research has been made possible, in part, by Accessibility Standards Canada.

\bibliographystyle{unsrt}  
\bibliography{references}

\end{document}